\documentclass[12pt]{article}
\usepackage{makeidx}
\usepackage{multirow}
\usepackage{multicol}

\usepackage[dvipsnames,svgnames,table]{xcolor}
\usepackage{graphicx}
\usepackage{epstopdf}
\usepackage{xcolor}
\usepackage{ulem}
\usepackage{hyperref}
\usepackage{amsmath}
\usepackage{amssymb}
\usepackage[toc,page]{appendix}
\author{sciences.univers@gmail.com}
\title{}
\usepackage[paperwidth=612pt,paperheight=792pt,top=70pt,right=70pt,bottom=70pt,left=70pt]{geometry}

\makeatletter
	{\par\setlength{\parindent}{#3}
	\setlength{\leftmargin}{#1}       \setlength{\rightmargin}{#2}%
	\advance\linewidth -\leftmargin       \advance\linewidth -\rightmargin%
	\advance\@totalleftmargin\leftmargin  \@setpar{{\@@par}}%
	\parshape 1\@totalleftmargin \linewidth\ignorespaces}{\par}%
\makeatother 
\DeclareUnicodeCharacter{2061}{X}

\begin{document}

\begin{center}
{\Large \textbf{Title} {The $\kappa$-model under test of the SPARC database}}
\end{center}

\begin{center}
{\Large G. Pascoli }
\end{center}

\begin{center}
{\Large Email: {pascoli@u-picardie.fr}}
\end{center}

\begin{center}
{\Large Facult\'{e} des sciences}
\end{center}

\begin{center}
{\Large D\'{e}partement de physique}
\end{center}

\begin{center}
{\Large Universit\'{e} de Picardie Jules Verne (UPJV)}
\end{center}

\begin{center}
{\Large 33 Rue Saint Leu,  Amiens, France}
\end{center}

\noindent
\Large{Abstract} {{
{Our main goal  is here  to make a comparative analysis between the well-known  MOND theory  and  a more recent model called  $\kappa$-model.  An additional   connection, between the $\kappa$-model and  two other novel  MOND-type theories:  Newtonian Fractional-Dimension Gravity (NFDG) and  Refracted Gravity (RG), is  likewise  presented.  All  these   models are   built
to overtake  the DM paradigm, or at least to strongly reduce the dark matter content. Whereas  they rely  on  different formalisms,
however,  all four  seem to  suggest that  the  universal parameter, $a_0$, appearing  in MOND theory could intrinsically  be correlated to either  the sole baryonic mean mass density (RG and $\kappa{}$-model) and/or to the dimension of the object  under consideration (NFDG and $\kappa{}$-model).   We could then  confer to  the  parameter $a_0$ a more flexible  status of multiscale parameter, as required to explain the dynamics together in  galaxies and in  galaxy clusters. Eventually, the conformal gravity theory  (CFT)  also seems  to have   some  remote link  with the $\kappa$-model, even though the first  one is an extension of    general  relativity, and the second one  is  Newtonian in essence.}} The $\kappa$-model has been tested on a small sample of spiral galaxies and in galaxy clusters. Now we test this model on a large sample of galaxies issued  from the SPARC database.}

\vspace{10pt}
\noindent
{\Large Keywords :   SPARC database, galaxy,  MOND,  Newtonian Fractional-Dimension Gravity, Refracted Gravity, $\kappa$-model, dark matter}

\section{Introduction}\label{sec1}

 As it is well known, all the studies conducted on 
galaxies and   galaxy clusters, lead to the seemingly firm conclusion that a
significant portion of the mass in the Universe  seems to be hidden from  the view 
of the observers. This invisible  (non-baryonic) matter is called  dark matter
(DM). It is true that DM is the more simple and economic hypothesis.  However, the
major problem with this paradigm is that the dark matter/baryonic (DM/B) mass ratio is incredibly 
huge, of the order of 6. It is not simply an addition of  a small quantity of 
missing  matter  to a dominant form of visible (baryonic) matter.  This is even 
the opposite, and the  baryonic component  eventually appears negligible  in the
Universe.  This situation seems to be surprising and even  rather unconfortable,
the visible sector  being explained by an undefined invisible sector about which we know nothing. De facto the explanation of the
flatness of the rotation spiral galaxy curves with DM is fully  indirect. A very
pertinent   parallel can be drawn  with the phlogistic  theory, a dominant theory in  nascent  chemistry  during the  18th century.  The  phlogistic  hypothesis was based  on the
existence of an illusive "substance" (the phlogiston) with indeterminate   properties  and thus  without real  physical support. The  theory of phlogiston was finally  disproved by the  French chemist  Antoine-Laurent Lavoisier through a series of experiments in the late {eighteenth} century.   
 Is it the   fate that awaits  DM?  At the present time  the existence of DM is inferred  only through  
gravitational effects.  A direct proof is missing from  both  an observational and an experimental
point of view.

The MACHOs (Massive Compact Halo Objects), possibly detected  through
gravitational microlensing in the Galactic halo, have been  ruled out as a dark
matter candidate [1]. Another  interesting 
interrogation is that if DM particles are really  existing, these particles can
very possibly decay. Strangely enough, X-ray space telescopes like  Chandra,
XMM-Newton, and Fermi have not observed any excess of DM decay [2].   Eventually a major  trouble  for DM is the tantamount  difficulties observing  the DM particles in the laboratory.   Large classes of candidates have
been suggested following highly speculative  theoretical  models, such as 
Hidden-Sector Dark Matter particles, completely neutral under Standard Model
forces, but interacting  through a new force; or still Ultra-Light Dark Matter
particles with predicted  masses from 10-22 eV to about a keV, and that can be
produced during inflation or phase transitions in the very early Universe [3]. However the existing dark matter experimental
programs are now more reasonably focused on weakly-interacting massive particles
(WIMPs)  [4, 5, 6].  Unfortunately the
conclusions of all these very costly studies are always negative. All direct
detections have come up empty.  The persisting  non-detection in space and in the
laboratory of DM  in spite of very intense efforts is rather discouraging.  A 
simple, but very frustating, conclusion  would be that if DM interacts uniquely 
gravitationally with baryonic matter and definitely not through one of the other
known three forces (the strong, weak  or electromagnetic forces), we might never
detect it. Another possibility is that DM interacts with itself and with 
baryonic matter but via an unknown (fifth) force.  In spite of all of that, DM  remains the leading explanation for the  dynamics of galaxies, very likely
 for its high flexibility adaptable to various situations (galaxies, galaxy
clusters and cosmology). This view can unfortunately persist for a very long time
because the DM paradigm seems to be unfalsifiable.  Yet a good question put by McGaugh [7] is however:  Is it a missing mass problem or rather an
acceleration-velocity discrepancy when observing the galaxies? Indeed the mass is an indirect data contrarily to the
acceleration which can be directly  measured. Di Paolo and coauthors [8]
  have remarked   that there exists a mysterious link between DM and
   the baryonic component. In fact this link is easily explained if 
  DM is a property of the baryonic mass itself.

Alternatively   a lot of  authors have  tried to circumvent  the trouble by 
exploring  other paths than DM. Without DM it is true that  the Newtonian theory
of gravity, and  even its basic relativistic version, i.e. the general
relativity,  seem to fail  on galactic scales.   The first model that  has been
developed in this sense  is the Modified Newtonian Dynamics or MOND [9, 10, 11].
Remarkably the basic idea of this model is thus as  simple and economic as  DM
concerning  the theoretical background.  The initial aim was to explain  the
flatness of the rotation velocity curves of the spiral galaxies uniquely with
help of  the observed baryonic matter. In MOND the second  law of
Newton ($ma=F$) is modified  in the very low regime of  acceleration
$a\leq{} a_0$, $a_0\sim{}1.2\ {10}^{-10\ }m\ s^{-2}$ being a universal
constant. MOND replaces the acceleration $a$ by $\frac{a^2}{a_0}$. Assuming then  a
test particle surrounding an attractive  mass $M$, with a circular orbit of
radius $r$ and with  $F=\frac{GM}{r^2}$, we have $\frac{a^2}{a_0}=\frac{GM}{r^2}$ or
$a=\frac{\sqrt{GMa_0}}{r}.$ For the velocity we directly obtain $v^2=ar=
\sqrt{GMa_0}$ $=Const$.  This leads  to the flatness of the observed rotation
curves of spiral galaxies but, much more, results in the Tully-Fisher law in a very natural manner [12].

Furthermore, MOND  is  sustained by the empirical Renzo's rule. The empirical
Renzo's rule~highlights the correspondence between detailed features observed in
the observational rotational curves of spiral galaxies and the same features seen
in their Newtonian counterparts [13]. This statement, that the observational rotation profiles 
seem to be a magnification of the Newtonian counterparts,   appears quite 
natural  when baryonic matter dominates the mass, but  not if DM is the dominant
form of matter. Another strong support for MOND, as seen above,   is the direct 
deduction,  within a calculation that takes just  a  few lines, of the Tully-Fisher
relation. These two  facts are difficult to explain within  the DM paradigm,
except in an ad hoc manner. Eventually MOND has predicted,  well in advance,
the profile of the rotation curves in the case of low surface brightness galaxies
(LSB), once again a feat not possible for DM [13]. However, the MOND phenomenology fails to explain the
dynamics of galaxy clusters. A natural remedy has been found by adopting a
multiscale approach [14]\footnote{{
{In the context of MOND, a multiscale approach adapts the parameter $a_0$ to the size of the system studied. However, in this case  the parameter $a_0$ is no longer universal.}}}.   In any way as for  gravitational lensing and 
cosmology, the classic modified-(gravity+inertia) MOND  in its initial  form  [9]   is not applicable. Various relativistic  versions  of MOND (RMOND)  have
been proposed  making  clear predictions regarding gravitational lensing and
cosmology. The latest in date is  that of 
Skordis and Z\l{}o\'{s}nik [15]. The latter
version of RMOND reproduces the  galactic and lensing phenomenology and also  the key
cosmological observables\footnote{{
{How the $\kappa$-model performs with lensing is presented
in the reference [16].}}}.

Another well known   modified-gravity  theory is the  covariant  
scalar-vector-tensor modified-gravity (MOG)  built by  J. Moffat  [17].  MOG is based on a $D = 4$ pseudo-Riemannian metric, a spin 1
vector field,  a corresponding second-rank skew field $B_{\mu \nu}$, and eventually three dynamical scalar
fields $G$ (the gravitational constant), $\omega$ and $\mu$.  The heavy price to be
paid is the addition  of extra vector and scalar fields to  the gravity field. 
On the other hand in  MOG  the gravitational constant $G$ is assumed to vary with
space and time.  Moreover the introduction of new fields means that  new particles are 
surreptitiously hypothesized.  We are not far from DM with its
elusive particles, even though MOG is much more subtle than DM because the
particles in MOG are virtual, and may not be  directly observable in the
laboratory.  MOG has been largely applied with some success  to  spiral galaxy
curves, to  galaxy clusters, to  gravitational lensing and eventually to
cosmology  [18]. RMOND and MOG are the two main models built to get rid of DM  fairly
efficiently. These two models are the only models that have been extensively
studied and involved in concrete comparisons with the observational data.
Unfortunately with  the relativistic extension  of MOND, or with its main
concurrent  MOG,  one moves away from   the beautiful simplicity of  the
Newtonian mechanics and even of   general relativity.  Let us note that RMOND
and MOG appear  very much alike. Thus the  major pitfall  of RMOND and MOG  is
the introduction of other subsidiary  extra scalar and vector fields that  have
not been tested in the laboratory.

A broad number of other models also exist, but they  have been
 more sporadically  applied to real situations. Conformal
gravity theories (CFT), which are compelling alternatives to general relativity
theory, have been  claimed to explain  the observed flat rotation profiles of
spiral galaxies, without invoking DM  or other exotic modifications of
gravity [19, 20].  Nevertheless  the extension of this type of models to the field
of  cosmology  appears  to be questionable.  Thus it seems that the Weyl  CFT\footnote{The Weyl  CFT  is  built by
replacing  the Einstein-Hilbert Lagrangian density, proportional to the Ricci
curvature scalar, by a quadratic contraction of the conformal Weyl
tensor.} cannot accurately describe the stated lensing
observations without again considering dark~matter [21].
Eventually another very different  way is  to conceive gravity, not as a 
conventional interaction, but rather  as an emergent property  [22]. In this case,
gravity is seen as an entropic force, i.e. closely related to thermodynamics. Testing this
hypothesis in the galaxy world is underway.

Are there other options to get rid of DM ?  We can answer this question   in 
the affirmative. Very recently and quasi-simultaneously,  a lot of new models  have
been proposed by following original, even though speculative,  ways [23, 24, 25, 26, 27]. These models sound similar, even though they use 
a different formalism. The aim is then to satisfy a principle of parsimony  in the
introduced concepts. It is indeed about three different strategies, but which
share  a number of  common features. All these ideas are new and still need
 deep understanding.

One  very aesthetic  strategy  is to assume that  spacetime is
multifractal in nature. This property  is revealed   in the most prominent  quantum
gravity theories in a natural manner  [28].
This concept of fractional-dimension space  applied  to Newtonian gravity  has
been suggested as an alternative to DM  [23, 29, 30, 31, 32]. In the latter work a connection has been established between the
Newtonian Fractional-Dimension Gravity (NFDG) with MOND. The
MOND acceleration constant $a_0$ can be related to a natural scale
length~$l_0$~in NFDG, i.e., $a_0\sim{}\frac{GM}{l_0^2}$, for any astrophysical
structure of mass~$M$, and the deep-MOND regime appears  in regions of space
where the dimension is reduced to $D\sim{}2$.

A second strategy is  Refracted Gravity  [33, 34].  Refracted Gravity mimics dark matter
by introducing  a  gravitational equivalent to a permittivity, seen as a
monotonic function of the local  mean volumetric mass density. This function is parametrized 
by  three coefficients which are free as in the case of DM, but which are
expected to be universal, contrarily to DM where the parameters are free and,
additionally  different for each galaxy. Once again even if  this second
strategy apparently  relies  on a very different  formalism  than NFDG,  both share
strong links with MOND.

We turn now to the third strategy, i.e. the $\kappa{}$-model. The aim of the $\kappa{}$-model is to  reflect on how the mean volumetric mass density (estimated at a very large
scale), surrounding a given observer, can modify his view of the Universe.

In   the framework of this model   [16, 25, 26, 27] it is hypothesized that it is the perception of the observer, modified
by his environment (the local mean volumetric mass density, calculated at a very large scale
around him),  that  creates the observed anomalies and also his proper experience 
of  gravity  (with today  the need to call for a hypothetical dark matter in
order to explain these anomalies). This  idea is  speculative, but it 
strongly resembles  the models for which we have given an overview above
[23, 24].  However one point of difference is that  the effects described in
the $\kappa{}$-model   are only apparent,  depending on the observer
(excepting the  spectroscopic velocities whose  measurement is universal, see [16] par. 2 eq. 10). It
is almost as if  we are looking at any object through a perfect, even though fictive,  optical device
(such as an aberration-free flat superlens\footnote{{
{ A superlens is a flat, lightweight option that can replace bulky traditional lenses and other components in optical systems. It is a lens that  goes  beyond the diffraction limit.}}}, but without being  aware of the
presence of this device (which obviously does not exist).   Clearly,   the object has  not changed but both its
apparent size and velocity can now appear magnified from  the point of view of a
distant observer. Admittedly both the inertia and the gravity seem to be modified
in the $\kappa{}$-model, but it is a pseudo-modified-gravity, it is not
of the same nature that a real modified-gravity as introduced, for instance  in
MOG or RMOND.  Moreover, in the $\kappa$-model  the gravitational constant (and the
speed of light), locally measured by any observer, are invariant. The
gravitational constant,  the speed of light and all physical  constants are
universal in the $\kappa{}$-model. To make variable  a constant in physics, in our case $G$ here, could  require   to  make variable  other constants (for instance the speed of light) with, may be, unpredictable consequences. Furthermore   no new field or exotic particles, 
undetected in the laboratory, are assumed in the $\kappa$-model. We think that it is a very  important
point that  obeys a principle of parsimony.  Eventually  even though the
$\kappa{}$-model is Newtonian in essence, its great advantage is that 
it can be naturally made relativistic. {
{A  first draft   of  what might  be  a relativistic version of the   $\kappa$-model is presented  in the reference [25]. However  in a galaxy the velocities $v$ of stars and gas are low compared to the speed of light $c$ (the ratio $\frac{v}{c}\sim  10^{-3}$), and the nonrelativistic approximation is sufficient, especially in the outskirts of galaxies where gravity is weak. The same  arguments also apply to  MOND}}.   For MOND a
notable relativistic version has however  been proposed [15].  Nevertheless the elegant
simplicity of the  initial  version of MOND has unfortunately  disappeared   in
the operation. At the Newtonian level the $\kappa$-effect is mimicked  by an apparent local scaling transformation applied  in an Euclidean space [16, 26].  In a Riemannian structure of space a local scaling transformation could be applied exactly  in the same manner.   Eventually let
us note  that the  multiscale approach already suggested  in  [14]   is directly included in the $\kappa{}$-model,
which assumes  that the larger the characteristic dimension (the scale) of a 
system, the weaker  the local mean volumetric mass density and the stronger  the magnification [16].

 In order to avoid any misunderstanding, three velocities are
defined in the $\kappa{}$-model: the Newtonian velocities, $v_{New}$,
which are  directly  calculable from the mean surface mass density profiles, but which
are virtual and not measurable, the radial velocities, which are given by
$v_{rad}\sim\ {\kappa{}}^{\frac{1}{2}}\ v_{New}$ (observationally
the universal spectroscopic velocities, $v_{spec}$) and the tangential velocities which are
given by $v_{tan}\sim\ \kappa{}\ v_{New}$ (observationally
the proper motions). Following a more mathematical approach within 
the formalism of bundles, the Newtonian velocities are "located" in the base
(not reachable)  and both the measurable radial and tangent velocities are
"located" in  a sheet, attached to a given observer in the bundle situated
"above" the base  [16]. The latter mathematical considerations will be shortly  developed  in  an
up-coming paper. We are only concerned here with the observational aspect.

\vspace{10pt}
{
{
{\raggedright
{The synoptic  table below summarizes the applicability domains  of the different models discussed  in this paper :}}
}}

{{
{\raggedright

\vspace{3pt} \noindent
\begin{tabular}{|p{100pt}|p{340pt}|}
\hline
\parbox{60pt}{\center
Model
} & \parbox{340pt}{\center
 
Main features
}
\vspace{2pt}
\\
\hline
\parbox{60pt}{\raggedright 
MOND
} & \parbox{340pt}{\raggedright
 
Very low acceleration   $a \lesssim a_0\sim{}{10}^{-10}\ m\ s^{-2}$
}
 \\
\hline
\parbox{60pt}{\raggedright 
 $\kappa$-model
} & \parbox{340pt}{\raggedright 

Very low mean  mass density  $\lesssim 0.15 \ M_\odot
{pc}^{-3}$

Geometry  of the matter distribution  (bulge, disk)

Compactness  (stars, gas)
}
 \\
\hline
\parbox{60pt}{\raggedright 
NFDG
} & \parbox{340pt}{\raggedright 
Variable dimension of the matter distribution, between $D=3$ (sphere) and $D=2$ (disk)
} \\
\hline
\parbox{60pt}{\raggedright 
RG
} & \parbox{340pt}{\raggedright 
Very low mean  mass density  $\ll{}0.17\ \ M_\odot\ {pc}^{-3}$

Geometry  of the matter distribution  (bulge,  disk)

} \\
\hline
\end{tabular}
\vspace{2pt}

}
}}

\section{Calculation details}\label{sec2}

In the SPARC catalogue [35] each galaxy is usually identified by  three independent
main components for the densities : the bulge labeled  $b$ in the following, the
stellar disk labeled  $d.st$ and the gaseous disk labeled $d.g$. This hierarchy is also preserved in the $\kappa$-model where both  the geometry and the relative values taken by  the mean densities (compact masses for the  stellar component,  or diffuse masses for  the gaseous component) are now playing   a new role by their implication in a magnification factor at a very large scale. A similar idea appears in the  NFDG theory, but  it is  the dimension of the matter distribution  that plays a major role. Let us note that the so-called $\kappa$-effect (a retranscription of the DM-like effect), said  in a practical way, is   a  "huge-volume-effect" and it  only occurs  at a very large scale; it is inexistent  at the solar system level (a bit like the quantum effects are fully imperceptible at the macroscopic level).  In the framework
of the  $\kappa{}$-model,   the relationship associating the corresponding (fictive)  Newtonian  velocities  to the measured 
spectroscopic velocity is  [16, 26, 27]

\begin{equation}
v_{spec}={\left(\frac{{\kappa{}}_{M_t}}{\kappa{}}\right)}^{\frac{1}{2}}\left[\frac{{\kappa{}}_{ref}}{{\kappa{}}_{Mst.b}} v_{b.st}^2
+\frac{{\kappa{}}_{ref}}{{\kappa{}}_{M_{d.st}}} v_{d.st}^2+\frac{{\kappa{}}_{ref}}{{\kappa{}}_{M_{d.g}}} v_{d.g}^2\right]^{\frac{1}{2}}
\end{equation}

\noindent where  each peculiar velocity  is  weighted by a $\kappa$-ratio. {
{The origin of the  $\kappa$-ratios results from the need to  take into account explicitly  both the   matter distribution dimension  (bulge or disk) and  the compactness of  this matter (stars or gas).}} In the $\kappa$-model all these  coefficients  are directly linked to
 the  mean  volumetric   mass densities  $\rho{}$ by a simple and universal relationship {
 {(ln denotes the natural logarithm)}}

\begin{equation}
\frac{{\kappa{}}_1}{{\kappa{}}_2}=1+ln\left[\frac{{\rho{}}_1}{{\rho{}}_2}\right]
\end{equation}

\noindent  with the necessary condition $\frac{{\rho{}}_1}{{\rho{}}_2}>1$. {
{The indexes "1,2" run on all the mentioned indexes.}}  The relation (2) is called universal in the sense that  this relation is valid   whatever the type of galaxies, and also for galaxy clusters [16, 26]. {
{In MOND the analog of   $\kappa$ is not a logarithmic function of the density, but  a rational function of the distance [9, 10, 11] (but both are sensibly  equivalent  in the case of an exponential distribution of matter)}}.  In 
relation (1) the indexes $ref$, $M_{t}$, $M_{b.st}$, $M_{d.st}$ and $M_{d.g}$ respectively designate the
reference value for the density, the maximum value $M$ of the total  density, $t$,
(stellar bulge, $b.st$, + stellar disk, $d.st$, + gaseous disk, $d.g$) estimated at the center of the galaxy, and the maximum value $M$ of each of the  independent
components, also  estimated at the center of the galaxy. The non-indexed coefficient
$\kappa{}$ is the  local one  (there where  resides the observer who feels  the
gravitational field). For practical purposes concerning  the disk components,  
the density $\rho{}$ can be expressed as a function of the observable surface
mass density  (indirectly obtained  from the brightness measurement), i.e. $\rho{}=\frac{\Sigma{}}{\delta{}}$
 with the thickness $\delta{}$, the latter quantity being here  assumed to be constant throughout a galaxy disk.  Apparently, the thickness of the disks seems to play a role in the 
$\kappa{}$-model,  very similarly to   what  is assumed  in the $NFDG$ model,
even though in the $NFDG$ model  it is the dimension of the mass distribution   that  intervenes
instead of the thickness [23, 30]. A variable  thickness  along a galactic
radius in the $\kappa{}$-model could have  a close connection  with the variable
dimension $D$ in the  $NFDG$ model. However it isn't as simple as it appears, and  we return to this issue in the following. The magnification coefficients of the
active mass composing  both   the stellar and   gaseous components  are  expressed
separately, resp. $\frac{{\kappa{}}_{ref}}{{\kappa{}}_{M_{b.st}}}$   
$\frac{{\kappa{}}_{ref}}{{\kappa{}}_{M_{d.st}}}$  and $\frac{{\kappa{}}_{ref
}}{\kappa{}_{M_{d.g}}}$ but  are still  calculated with the same universal  
relationship (2).  When  the mean surface mass density is larger than $500\ M_\odot\ pc^{-2}$, a saturation effect  appears for   $\frac{{\kappa{}}_{ref}}{{\kappa{}}_{M}}$, and then  in all the cases we  put this factor  invariably equal to $0.45$, as provided  by the relation (2). However, in a few rare situations, especially  for  galaxies  with a big bulge, and in order to adequately fit the observational profiles in the inner regions, we should  adjust the factor $\frac{{\kappa{}}_{ref}}{{\kappa{}}_{M_{b.st}}}$  to a value between 0.45  and 1. An explanation to this statement  is that, in fact,  the relation (2) is valid for a thin disk, but not for a 3D bulge. At this level a clear reference to the  NFDG model where the dimension  plays a major role can be noticed. Another explanation is that  the  bulge of a spiral galaxy  is a very complex system where the stellar  orbits are randomly oriented. Then we  know that  a severe velocity dispersion, larger than $\sim  50 \ km/s$  a few kpc from the center,  can strongly  affect the extraction of the pure rotation velocity {
{(see for instance  the reference  [36] for the Milky Way)}}. The part  of the cylindrical rotational support  in  the inner regions of a spiral galaxy  is  generally difficult to estimate when the bulge is dominant.

A fundamental question is: how many free parameters  are used in
the $\kappa{}$-model ? We know that in physics,   the
fewer parameters, the better the model.  Yet by consulting   relation (1), we
see immediately  that four parameters (the $\kappa{}$-ratios) appear. Following the  parsimony principle  it is not a "good" model.  In fact,
once the density in the bulge, in the stellar and  gaseous  disks is
provided,  the $\kappa{}$-ratios, which are directly  issued from  observational data, are automatically  determined, there  is no longer   free
 parameters and the   $\kappa{}$-model eventually  becomes parameter-free   (the only  parameters being, as usually, the  observables, i.e.  the
surface brightness, the inclination and the distance, even though   unfortunately not very well known in some cases).  This is in strong
contrast with DM where two or three free external, and arbitrarily chosen, parameters are introduced  to just
obtain the expected results. However given that  the $\kappa{}$-ratios  are
dependent  on the  densities, the parameters in the $\kappa{}$-model    can now  vary
from one object to another, and this confers some flexibility to  the model with no violation of  the parsimony principle.  For
instance, the $\kappa{}$-model  has  been applied with success  to the physics of galaxy
clusters [16].  The mean mass density  in a galaxy cluster  is lower by three orders of
magnitude  compared to  the mean mass  density in a galaxy. The
$\kappa{}$-model is then  naturally  a multiscale model (or density-dependent scale model),  like the one  proposed
in   [14]    for the application of MOND to  galaxy clusters. The
difference is that in the  $\kappa{}$-model the scaling is not imposed, but appears  
in  essence,  taking  its origin in  the hierarchy of the   mean mass densities.  By
contrast, MOND [9, 10, 11]  with just  one universal parameter  or even the
Refracted Gravity [33, 34]  with three universal parameters   seems to be too rigid.
On the other hand,   the $\kappa{}$-model can naturally be made
relativistic [25],  making possible  its extension to  cosmology, especially to the
 analysis of the fluctuation density  in the CMB. In this case
it is the density  anisotropies to  mean density  ratio which intervene in the
relationship (2). The latter  very important  topic will be examined in a next
paper.

 Now, if we want  to compare the $\kappa{}$-model and MOND, we must 
define  a reference point  for the mean mass  density ${\rho{}}_{ref}$. Unfortunately, this
quantity is only  indirectly known by the ratio  $\frac{\Sigma{}}{\delta{}}$ 
(surface mass density, $\Sigma{}$,  over the disk thickness, $\delta{},$  in a spiral
galaxy such as the Milky Way).   The link between the acceleration parameter 
$a_0\sim{}1.2\ {10}^{-10}\ m\ s^{-2}$ of MOND and the reference surface density 
${\Sigma{}}_{ref}$ is

\begin{equation}
{\Sigma{}}_{ref}=\frac{a_0}{2\pi{}G}=152\ M_\odot\ {pc}^{-2}
\end{equation}

Let us note that this value is relatively close to  the  galactic
surface mass density  estimated in the solar region ($\sim   70 \ M_\odot\ {pc}^{-2}$) {
{(in comparison with  the high range of surface densities seen  in a disk galaxy, varying  from $\sim 1000\ M_\odot\ {pc}^{-2}$  in the inner regions,  $1\ kpc$ from the center, to $\sim 1\ M_\odot\ {pc}^{-2}$ in the outskirts,  $20\ kpc$ from the centre)}}.  Taking into account  the fact that the range of mean mass  densities is very extended in the Universe, this  appears indeed  very  odd if we see the parameter $a_0$ as a cosmological parameter; because in this case we must assume that our situation  in the Universe is privileged. In reality, the $\kappa$-model easily explains this rather  strange coincidence. We chose this reference taking  into account our position in the galaxy, but which has nothing special.  Another observer, located elsewhere, will take his own reference. The  relation (1), giving a universal result in the framework of the $\kappa$-model, his measurements would  lead  exactly to the same results for the spectroscopic velocities as ours, even though with his proper local  reference for the mean mass density.

 Figure 1 displays a panel of  velocity profiles for MOND and the
$\kappa{}$-model,  in the schematic situation of  a disk of matter where the mean surface mass density
varies  exponentially  (the thickness is assumed to be constant
following the radius $r$). The comparison between MOND and the $\kappa{}$-model
shows  that the logarithmic relation (2) is a very good choice. In MOND the
function $\mu{}\left(r\right)$  plays a very similar  role (see eqs. (7) and (8) of the reference [13]), {
{even though in MOND  $\mu{}\left(r\right)$ is not a logarithmic function,  but a simple rational function.}} We can note that the $\kappa$-effect (or MOND-effect)  plays a decreasing role, when going from low mass surface density (LSB galaxies) toward  high mass surface density (HSB galaxies), as confirmed by the observations. This finding, naturally explained with MOND or  the $\kappa$-model, remains unexplained in DM.  A
difference between MOND and the $\kappa$-model  is however  perceptible for the schematic representation  of a
so-called super spiral [37]. For a
high surface density (${\Sigma{}}_M \sim 10\ 000\ M_\odot\ {pc}^{-2}$, {
{see for instance the reference  [38]}})   the
$\kappa{}$-model curve   is located  more than $200\  km/s$ above  the MOND
profile for the terminal velocity (fig. 1d).

 \begin{center}
\includegraphics[height=460pt,width=250pt] {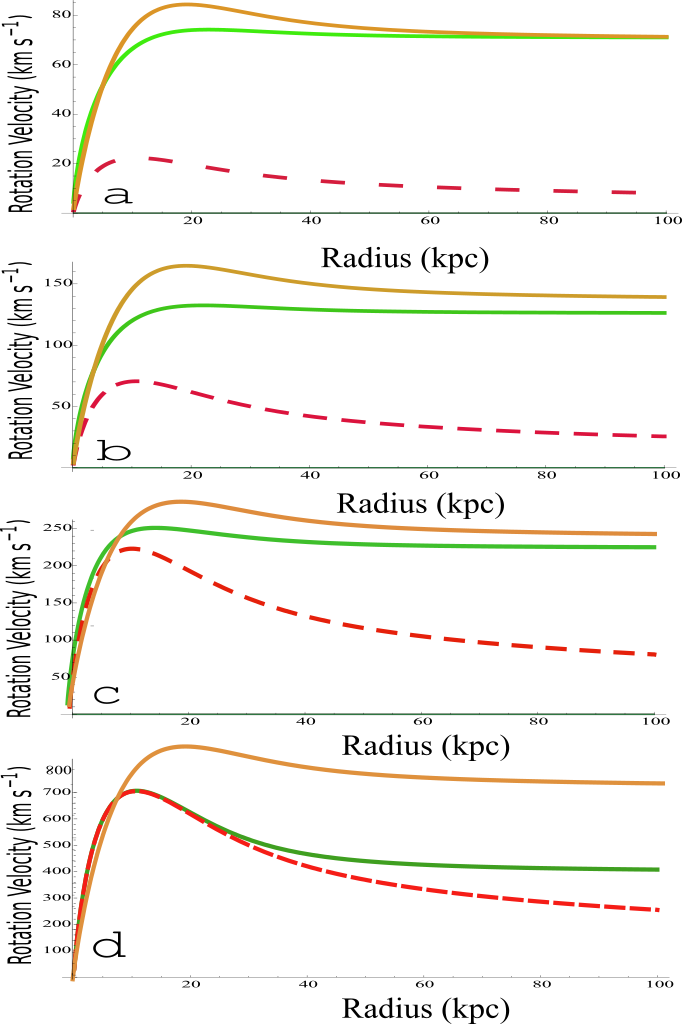}
\end{center}

Figure 1 : {
{Schematic  galaxy velocity curves  fitted with a simple exponential surface
mass density profile of uniform thickness}} (in the approximation of the thin disk). MOND is the green line, and the
$\kappa{}$-model is the amber line. The dashed red line is the Newtonian curve
(baryons). The reference value for the surface density is 
${\Sigma{}}_{ref}=\frac{a_0}{2\pi{}G}=152\ M_\odot\ {pc}^{-2}$. $a.\ {\
\Sigma{}}_M=10,\ $    $b.\ {\ \Sigma{}}_M=100,\ \ \ $   $c.\ {\
\Sigma{}}_M=1000,\ $   $d.\ \ {\Sigma{}}_M=10\ 000\ $    $M_\odot\ {pc}^{-2}.$

\vspace{10pt}
In the more concrete cases, the situation is obviously different  from  the previous 
trial examples with a simple  exponential disk. In reality,   we
encounter in the SPARC catalog a number of situations where it is not possible
to fit the mean surface mass density of both   thin stellar, ${\Sigma{}}_{d.st}$, and gaseous,
${\Sigma{}}_{d.g}$, disks by just adopting   a simple  exponential fit.  In these situations,
we had to add  to the exponential component  one, or sometimes  two, decentered
gaussian components. The velocity curve $v_d(r)$ is then deduced from  the
self-evident, more general formula (still assuming  an axisymmetric disk) valid for one component (stars or gas)

\begin{equation}
\frac{v_d^2(r)}{r}=G\int_{\Omega_\infty} dxdy\frac{\Sigma{}\left(\sqrt{x^2+y^2}\right)(r-x)}{\kappa(r){\left[{(r-x)}^2+y^2\right]}^{\frac{3}{2}}}
\end{equation}

{
{As a first step  the  operational  method consists  to  fit the   Newtonian velocities available from the SPARC catalog,  for  the  distributions of stars and gas, taken individually, and for each galaxy. In simple terms we fit  the dashed-red  (stars) and dashed-green (gas) curves of  Figure A. In this case   the  relationship (4) is applied with  $\kappa(r)\equiv 1$ (this is the usual Newtonian level). Secondly,  the same relationship (4) is again used, but incorporating this time the coefficient  $\kappa(r)$ that depends on the  volumetric mass density (eq. 2). This second step automatically  provides  the corresponding  $\kappa$-model curves. The great  benefit  of the method is that all parameters are internal to the theory, and   supported   by  the sole  observational data, essentially the baryonic mass density.  There is no  arbitrary  parameter such as  the ad hoc DM/B ratio in DM.}} 

 Eventually,  when a bulge is present, a de
Vaucouleurs formula  [39] is used to fit the surface mass density of the bulge. Two other 
parameters, intervening in the $\kappa$-model,  are still  the thickness (scale height) of the stellar (thick) disk, ${\delta{}}_{st}$,
 and that of the gaseous (thin) disk, ${\delta{}}_{g}$.    For all the galaxies under study (SPARC
catalog), these 
parameters  have been  taken to be  equal to the reference values estimated for the Milky Way   in
the vicinity of the Sun, respectively ${\delta{}}_{\odot.st}$   and
${\delta{}}_{\odot.g} \sim  0.5\ {\delta{}}_{\odot.st}$. Given  that the galaxies are
diversely  oriented with any inclination angle, these parameters  are difficult to
estimate and certainly variable along a galactic radius. Our analysis of the
SPARC galaxies seems to indicate a neat trend where the thickness decreases when
going from the core regions to the outskirts in the  flattened galaxies.

Globally, for a mean orientation of $45^\circ{}$  the thickness along
the line of view  is increased by a factor of $\sqrt{2}.$ In this case the rotation
profiles provided by the  model  have  to be magnified by a few percent. In
fact, the logarithmic function flattens the density ratios in  relation (2) and
the influence of the variation of the thickness has a strongly reduced, even
though not negligible,  impact on the corresponding  $\kappa{}$-ratios (of the
order of $10\%$ for a  thickness variation  of a factor 2. For orientations
larger than $45^\circ{}$ the magnification can obviously be much larger than $10\%$). Let us note that
in other models  where the density is assumed to play  a role, for instance in 
[24, 30],  the  conclusions  should  be very similar when applied to a large sample of galaxies, such as the SPARC database.  The thickness
 of various types of spiral galaxies  has been estimated by
different methods [40, 41, 42].  For
irregular dwarf galaxies, the situation appears relatively confusing, but the latter
category  can exhibit  quasi-round galaxies with a high mean thickness  [40]. The measurement  of the   thickness seems  to give values   of the order
of  ${\delta{}}_{\odot.st}$  or  ${\delta{}}_{\odot.g}$   to within  a
multiplicative in the range 0.25 (in the outer  regions) to 4 (in the inner regions toward the bulge if existing),  compared to the reference values, independent of the size of the galaxy (with a few exceptions for the very small galaxies, where smaller values for the thickness are favored). Thus, a positive point is that  an estimate of the thickness can be reached in the framework of the $\kappa$-model. However in figure A, for all the galaxies and for the sake of homogeneity, the thickness has been   taken as invariable throughout the stellar and gaseous disks. The  corresponding values are indicated in each individual figure. Taking into account a variable thickness would make it possible to obtain better profiles. A work that remains to be done.

Additionally, let us specify that   the  observations rather provide non-monotonous  galactic rotation
profiles.     Nevertheless it is illusory to try to perfectly fit the rotational
curves with  their  delicate    patterns  of bumps and wiggles. Very likely, these 
patterns  are caused by the presence  of spiral arms or portions of rings, a variable thickness or inclination,  not taken into account  by
assuming smoothed axisymmetric and monotonous  density profiles.  Even DM with   two or even three external 
parameters cannot make that\footnote{{
{A list of DM methods with two or three ad hoc parameters is presented  in the references [43, 44].}}}.  One of the better DM methods,   built on the  Einasto  profiles  with  three ad hoc parameters  in the fits, is discussed  in reference [45]. We can see that the fine details cannot be adequately fitted (see, for instance, NGC6015, NGC 7793,  NGC3726, IC4202, NGC0289, UGC06787, etc).  In any case  a lot of physical parameters are very poorly known: the inclination of the galaxy   (moreover, very likely variable along the
galactic radius), the mass-to-light  ratio, the thickness along the line of
sight, the distance, etc. We must add that the observational profiles can
 substantially differ in some cases from one author to another, sometimes by more than $20\ km/s$. We can compare two different catalogs, for instance, that of   Sofue  [46] versus SPARC [35], when the rotation curve for the same galaxy is presented (see especially NGC 2903 where a discrepancy of $40\ km/s$ can be notified).  Even for the Milky Way, in the vicinity of the Sun, divergences also exist [47].  Let us note that  the  DM paradigm could,  however,  be made  in agreement with any  inclination by adequately adjusting the DM/B rate!   A contrario both MOND and the $\kappa{}$-model apparently fail if
the inclination is not accurately estimated [26].  An example where the inclination factor  can sometimes play  an important
role in the determination of the rotation velocity  profiles is given in [48]. In the latter paper it is shown  that the inclination can
vary  by $20^\circ{}$ following the authors, eventually favoring a model rather
than another one. Eventually, we can say that, unfortunately, the determination of the inclination is not the sole trouble. Additionally   the gas and the stars in a galaxy, following their types, 
 do not rotate in the same manner,   the velocities are not circular, the galaxy disks
are not symmetric, etc.  The multiple consequences on the observational profiles are difficult to estimate.  This is  why  various  observational techniques  can lead to different profiles for the same galaxy. 

In spite of these difficulties, and in order to make a valuable comparative analysis between different theoretical models, the idea is to use a  same set of  extended data. For instance, 
the SPARC catalog seems in this case necessary.  This catalog  gathers   a large, and homogeneous, sample of 
rotation profiles. A very good point of the SPARC database is that it represents  a
uniform estimate of the surface densities of galaxies,  starting from  Spitzer
near-infrared data [35].  Then our procedure  as to starting from mean fits of the
Newtonian curves, and then mean fits for the observed rotational curves can be
deduced.  In some cases, the DM fits seem  to be  much more  impressive  [44, 45], but a major  drawback for a physical model is
that the DM technique of  fitting is not at all predictive.  Then, starting from
any Newtonian curve (even false), we can build any "good" predicted
 profiles, obviously by adding the "good" rate of DM. Admittedly, MOND
and $\kappa{}$-model profiles are  generally of lesser quality, but in most 
cases, both of them produce a good trend for the fits compared to the
observational rotation curves. Let us specify again that the latter ones, empaired  by various  biases,  are also not  perfect either.

\section{Results}\label{sec3}

\subsection{MOND versus $\kappa$-model}

 The results of the calculations for the individual galaxies in the  SPARC
catalog  are  collected in Figure A in the Appendix. The galaxies have been classified  in  alphabetical order to facilitate the research. For the disks, it is assumed that the  thickness is constant along a radius of the galaxy. In most  cases, the thickness has been taken to be equal to the corresponding reference values taken  at the Sun position in the Milky Way,  for both the stellar, let $\delta_{\odot.st}$, and  gaseous disk, let $\delta_{\odot.g}$. In view of the results, a first general remark is that  the $\kappa{}$-model is clearly as predictive as MOND\footnote{The MOND profiles have  been obtained with the formula

\begin{equation}
v_{MOND}^2=v_{bew}^2{\left[\frac{1}{2}+\frac{1}{2}{\left[1+4{\left(\frac{a_0}{g_{new}}\right)}^2\right]}^{\frac{1}{2}}\right]}^{\frac{1}{2}}
\end{equation}

where $v_{new}$ and $g_{new}$ are respectively the Newtonian velocity and acceleration.}.  For  both models,
the results statistically  deviate   by less than $10\%$    as for the
prediction of the  terminal velocities (Figure 2).

\begin{center}
\includegraphics[height=160pt, width=210pt]{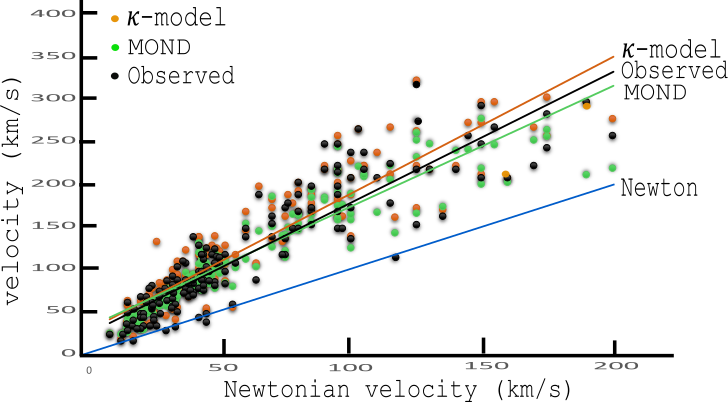}
\end{center}

 Figure 2  The terminal velocities for the sample of galaxies (SPARC data base).  MOND is in green, the $\kappa$-model is in amber, and the
observed velocities, provided by the SPARC database, are represented  in black. Linear regression lines are also represented. 

\vspace{10pt}
Examining the individual cases, by browsing   figure A,  we can see that the $\kappa$-model  leads
to predictions similar to the MOND phenomenology, even though in some cases,  the profiles are not quite identical. Moreover, a comparison with the observational profiles shows that   the  predicted curves for  both  MOND and the $\kappa$-model don’t perfectly match the observations\footnote{We can remark that whereas  some authors affirm that the MOND fits are fairly good [47, 49], on the other hand,   other authors, who  rather seem to defend the DM paradigm, conclude that the MOND fits are  bad in a large percentage of analyzed individual cases [50, 51]. It is true that  MOND gives fairly acceptable fits in a large number of  cases and is less good in other cases. This situation can easily be explained if we admit that the mass-to-light ratios, the inclinations, and the distances are poorly  known. We can then postulate that MOND systematically provides   "perfect" fits and can then  predict the inclinations and distances. A contrario  DM fits  are apparently  better but,  ironically enough, even if we  choose  a bad inclination or an erroneous  distance.}.  Reporting to figure A we see that  the theoretical curves predicted by MOND or the $\kappa$-model can be indifferently located slightly above or below the observational curves. However, there are remedies for this.  First, in the outer regions,  the predicted curve is quite  often  located above the observational one. In MOND   the bias 
 can then be  corrected by EFE (External Field Effect) [52].  Likewise, this bias  could  be corrected by a diminushing thickness
of the disks (at constant surface density $\Sigma{}$)  in 
the $\kappa$-model.    A contrario,  there exist a number of cases where the predicted curve is located  below the observational one, especially in the innermost regions of galaxies  (for instance, in the more striking cases:  F563-V2, F568-1, F568-V1, F571-8, F579-V1,  F583-1, NGC 2915, NGC 3992, NGC 5907, NGC 5985, NGC 6674, UGC 00128,  UGC 00731, UGC 02259,  UGC 6446, UGC 06667, UGC 07399, UGC 7490, and  UGC 8286). It is interesting to  note that the magnitude of this   bias is very similar  in both  MOND and in the $\kappa$-model for a  given galaxy. Very possibly,  a modification of the inclination  in  MOND (and also  in the  $\kappa$-model)   could partially  remove the discrepancy in the inner regions  for these galaxies. As demonstrated in [48] the modification of the inclination can substantially modify the profile of the rotation curve. However this effect  appears rather systematic throughout  figure  A in the inner regions. In other words,   the measurement of the inclination would be systematically biased in the inner of spiral galaxies and, strangely enough, always in the same direction. This hypothesis is hardly acceptable. The fact that  the MOND curve is located below the observational one results from the fact that the acceleration $a$ is equal to or larger than  the critical value $a_0$ in the inner regions.  In this case, we are in a domain   where the Newtonian regime is still supposed to be valid.  To save MOND, we can then assume that the parameter $a_0$ is larger in the inner region, but then this parameter is no longer universal. Another solution is that the baryonic mass-to-light ratios are largely underestimated (by a factor 2) in the inner regions of  the quoted galaxies.  Eventually, a more credible explanation is to imagine that some non-exotic form of DM  exists in the innermost regions of galaxies. We could invoke, for instance,  a  neutrino species  with  mass $\sim\  eV$ (but in acceptable  quantity with $DM/B \sim  2$). A  very similar idea has been assumed  for the inner regions of  galaxy clusters  [53]. This hypothesis appears admittedly reasonable; however, the $\kappa$-model can propose  another natural solution.   In the $\kappa$-model, the leading  role is not  held   by a fixed parameter, i.e. the acceleration $a_0$, but by the  mean volumetric mass density.  This hypothesis makes  the  $\kappa$-model  more flexible than MOND. Figure A displays the results under the reductive hypothesis of  a constant thickness throughout the galactic disks. However  an increase in disk thickness (at constant mean mass surface density) in the inner regions of spiral galaxies  could help  to lessen  the discrepancy. An interesting conclusion is that the  $\kappa$-model could hence  help to obtain  an estimation of the mean thickness, a parameter difficult to derive from the observations. In any case in the cases mentioned above, even if the $\kappa$-model gives imperfect fits in the innermost regions of these galaxies, we can see that the terminal velocities are correctly predicted. A simple response to these statements  is that  the empirical relationship (2) is very well adapted to a thin disk, but far less applicable to a thick disk or a 3D bulge. {
{ For the    galaxies listed just above,  where a  discrepancy  exists between  the  $\kappa$-model (or equivalently MOND) and  the observational data}},    a comparison with  DM profiles  with  two or three (ad hoc) parameters (as in reference [44]) appears  very interesting.   Examining the cases displayed in figure 6.10  of [44], we can see that the underlined  discrepancy also persists in some of the  cases,  even though slightly  lessened  (see, for instance, the rotation profiles  for F568-1, F579-V1, NGC 2915,  NGC 5907 and  NGC 6674).  F571-8 is a pathological  example where MOND, the $\kappa$-model and DM yet  provide very  similar profiles, but paradoxically enough far from the observational one in the outer regions.  The three  theoretical profiles, even though   very similar,  are located  $50\  km/s$ below the observational profile in the outer regions.   In any  way, we know that  trying   to  predict the rotation velocity curves with a better statistical precision
than $10\%$ (and even, in some pathological cases, the incertitude can rise to $20\%$)  appears unwarranted, considering the dispersion in the observational
data  coming from various sources.  That matter aside, in the framework of the $\kappa$-model, at least we have  a fairly good 
estimation  of the terminal velocities (Fig. 2 and see also  figure A for the individual cases), a conclusion that  cannot be reached  by
the ad hoc  DM methodology. Obviously, the flexibility of the $\kappa{}$-model  by
taking  into account a variable  thickness  in
the stellar and gaseous disks would allow to fix  the residual discrepancy between the theoretical  curves and the observational ones. In the same vein, this statement is
rather   attractive  because it implies that   the $\kappa{}-$model   could
predict  the variation of the thickness in   spiral galaxies along a radius. This data is  indeed difficult to obtain by  sole observation.

\subsection{Newtonian Fractional-Dimension Gravity}

{
{Newtonian Fractional-Dimension Gravity (NFDG) is  an extension of the laws of Newtonian gravitation to
lower dimensional spaces, including those with non-integer, "fractional" dimension (for a general introduction see [23]).  NFDG is based on a
generalization of the gravitational Gauss’s law, replacing standard space integration over $\mathbb{R}^3$  with an appropriate Hausdorff
measure over the space, which was related to Weyl’s fractional integrals.  As for MOND or  $\kappa$-model, the goal of NFDG is to describe galactic dynamics without using the controversial DM component. A quick review of NFDG is presented in the reference  [30].  NFDG was introduced heuristically by extending Gauss’s law for gravitation to a lower dimensional
space-time $D + 1$, where $D \leq 3$ can be a non-integer space dimension. A scale length, $l_0$, is needed to ensure
the dimensional correctness of all expressions for $D \neq 3$. Let us note that NFDG does not imply a change in the tri-dimensionality of space in galaxies, but rather  the local Hausdorff dimension $D \neq 3$ is associated to  the matter distribution (bulge or disk). In this sense there is an   analogy with the $\kappa$-model, where the $\kappa$ ratios   (eq. 1)  are assumed to be dependent on  both  the dimension of the matter distribution (bulge or disk), and  also  the compactness of matter (stars or gas)}}.

In [29] Varieschi discusses in depth the case of  NGC 6503. For NFDG with a dimension $D=2$, the theoretical curve is slightly above the observational one and is remarkably flat (see Fig. 6  of [29]). However, assuming  that NGC 6503 behaves as a fractal medium, with a variable fractional dimension, NFDG  can produce a curve with a perfect superimposition with the observational one. Reporting now  to Fig.A for this galaxy we can see that  both  the  MOND and  $\kappa$-model curves  are slightly  below the observational curve in the inner regions and slightly above in the outskirts.  In the $\kappa$-model framework, the statement of variable fractional dimension could be  re-interpreted as a variable thickness of the disk. In the case of NGC 6503 for instance,  an increase in the thickness in the inner regions (thick disk) and, concomitantly,  a decrease in the thickness in the outer regions (thin disk),  could also lead to an improved  profile, such as in NFDG theory. {
{In  [30] the same author applies  his analysis to  other  rotationally supported galaxies : NGC 7814 and NGC 3741,  for  which  very good NFDG fits are supplied}}.  If we consider these  galaxies, MOND and the $\kappa$-model provide a fairly good value for the terminal velocity. However, a same bias is perceptible for the inner velocities (the predicted curve is below the observational one).   This bias is not present on the NFDG profiles, which perfectly fit the corresponding observational curves, even with their humps and wiggles. However this perfect fit results from   the fact that  the  NFDG theory imposes on  the fractional dimension  to vary in "an appropriate manner"  along a galactic radius, in order to obtain a "good" profile. Nevertheless, the positive point of this  procedure   is  that  NFDG can thus be  predictive for  variable dimension. Once again,  the $\kappa$-model can correct the mentioned  bias by invoking a variable thickness.  In the framework of this model,   a volume-effect, i.e.   the influence of the mean volumetric  mass density surrounding a given observer, and  estimated at a very large scale,   is playing a similar    role to that of the dimension in NFDG.  Yet Varieschi underlines that the variable dimension  $D$ should be interpreted as the dimension
of the matter distribution of the galactic structure  and definitely not at all as the local space dimension that an
observer would measure at a specific galactic location. In any event, the link between a variable  dimension in NFDG theory and a variable thickness in the $\kappa$-model could be more subtle, and should be reconsidered in more depth. Furthemore, examining relation (1), we can see that the coefficients $\kappa$ for the bulge and the stellar and gaseous  disks are   different. For the bulge and the disk, the dimensions  are admittedly  different, but   what about for the stellar versus  gas components ?    All these {
{questions}}   deserve   further examination.

It will be very interesting for comparison with the $\kappa$-model that the NFDG theory be applied  to a larger  sample of galaxies,  for instance, the totality of the galaxies of the SPARC database. A particular attention must then be paid to  the following cases : F563-V2, F568-1, F568-V1, F571-8, F579-V1,  F583-1, NGC 2915, NGC 3992, NGC 5907, NGC 5985, NGC 6674, UGC 00128,  UGC 00731, UGC 02259,  UGC 6446, UGC 06667, UGC 07399, UGC 7490, and UGC 8286, for which  both  MOND and $\kappa$-model substantially differ  from the observational profiles in the innermost regions, while however   providing  fairly good estimates in the outskirts  of these galaxies (the terminal velocities).

\subsection{Refracted Gravity}

Along with the  NFDG model, another new  classical gravity modified theory is the so-called  Refracted Gravity (RG)  [24, 33, 34]. RG can be reformulated as a scalar-tensor theory [34].  {
{RG mimics DM with a  gravitational permittivity (a kind of  variable gravitational "constant" $G$),  and  that boosts the gravitational field in low-density
environments. In RG the link between the volumetric mass density $\rho$ and the gravitational permittivity $\epsilon$ is expressed by using the  relationship }}

{
{
\begin{equation}
\epsilon(\rho)=\epsilon_0+\frac{(1-\epsilon_0)}{2}\left\{tanh[ln(\frac{\rho}{\rho_c})^Q]+1\right\}
\end{equation}
}}

{
{where $\epsilon_0$, $Q$ and $\rho_c$ are three free parameters. The formula (6) is  an arbitrary monotonic function of the volumetric mass density with the asymptotic
limits $\epsilon(\rho)=1$ for $\rho >> \rho_c$ and $\epsilon(\rho)=\epsilon_0$ for $\rho << \rho_c$.  This formula is the  equivalent in RG  of  the  relation (2) in
the $\kappa$-model. This permittivity also  shares  a very strong analogy with  the function $\mu$ in MOND [13],  or still  the function $\kappa$  in the  $\kappa$-model [16, 26]. However $\epsilon$ is supported by three   universal parameters, instead of just one, for instance as in MOND ($a_0$). Thus  RG seems, at first sight,  to be less economic than MOND, but its great  interest is that it is now a multiscale version of MOND. In this sense, the objective  of  RG is very  similar to that proposed  by the $\kappa$-model; but with an  essential difference :  the $\kappa$-model model    uses exclusively  internal parameters (i.e. the mean volumetric mass densities of the bulge, stellar and gaseous disk components) and no free  external parameters. Then, by contrast  in  RG   the  three arbitrary parameters still  need   to be  obtained  through  a long statistical  analysis of the  observational data [24]. RG has been applied to both flattened galaxies  [24] and a small number of elliptical   galaxies  [33].  The  results presented in [24] rely on setting the three free parameters for each individual galaxy. However, the authors show  that the variations  of these parameters from galaxy to galaxy can, in principle, be ascribed to statistical fluctuations. Then the authors adopt an approximate procedure to estimate a single series of parameters that may properly describe the kinematics of the entire sample of galaxies,  They eventually conclude that the gravitational permittivity is indeed a universal function. }} Unfortunately, a direct, and yet  fruitful, comparison between RG  and the $\kappa$-model is difficult because the galaxies under consideration  are not issued from the same  catalog. However, a close examination of the results displayed  in [24] leads to the firm conclusion that the  fits of the rotation profiles are of similar quality to those  produced by MOND and the $\kappa$-model. 

 \subsection{Conformal Gravity}

Eventually, a comparison with conformal gravity can also be proven worthwhile. In the Conformal Gravity (CFT) [20] two universal parameters are introduced, setting apart the usual  observational data, i.e. the luminosity and the $M/L$ ratio, the distance, and the inclination, common to any model. {
{The first parameter  ($\gamma*$) is related to  the local geometry, while the second parameter ($\gamma_0$)  describes  the global geometry due to all the other galaxies in the Universe.}}  These two  parameters are statistically derived from  the observational rotation  curves of a chosen sample of 104 galaxies (this sample is limited to the galaxies whose   mass density is  fittable by a simple  exponential thin disk). By comparison,  we recall that in the $\kappa$-model the coefficients $\kappa$ are calculated from the mean mass density profiles attached to each galaxy. However,  it is very difficult to decide which model is the best.   Statistically, MOND, the $\kappa$-model and CFT provide equivalent results as for the proximity of the theoretical curves  to the observational ones.  We can examine the fit through the individual cases presented in [20].  For NGC 1003, NGC 3972, NGC 5585, and UGC 7089,   MOND and  $\kappa$-model fit is better than the CFT fit. For NGC 2903, UGC 5005,and  UGC 5999 the  fits are equivalent,   For  NGC 4100 the  CFT fit is better  than the MOND and $\kappa$-model fits, etc. Some cases  are  favorable to  MOND or to   the $\kappa$-model while in other cases the CFT is better.  At the present time, this situation is very embarrassing because each author can validly support his own  model against that of others through a judicious choice  of the data. It is for that reason that the models have to be compared on a very large sample of galaxies such as the SPARC database, and not on a very small sample of a few galaxies.

\section{Conclusion}\label{sec4}

 This paper is a discussion  on the capacity of  a number of MOND-type  models [16, 23, 24] and a CFT-based  model [20], which have been recently proposed,  to understand the dynamics of a large variety of flattened  galaxies. Admittedly, these models do not provide very perfect fits (except may be  NFDG that possesses a flexible dimension associated to the mass distribution),
but,  nonetheless, they produce  fairly  predictive mean
rotational curves.  It is nonetheless true that DM   can indeed
lead to   better  fits with two  or  still three [44, 45] 
parameters, but unfortunately,  these parameters are freely adjusted to each galaxy. This implies that  by starting from any Newtonian profile, even one  strongly empaired by  various biases, we can derive  a "very good"  fit for  any given  observational rotation profile\footnote{Once again in the framework of the $\kappa$-model starting from the Newtonian curves, we generate a predictive profile  for the observational one in an univocal manner. The baryonic mean  mass  density  alone is the conductor of the situation.}.  MOND and the $\kappa$-model are  at least  falsifiable and upgradable, while  DM definitively not. For a physicist, the choice is quickly made. 
With no confirmation by experimental methods, DM unfortunately has   very limited
scientific significance.   Obviously this  conclusion would drastically  change
if,  one day, we discover the signature of DM in the laboratory. We can
always expect  it  over the next few years. Even though  obviously  the $\kappa$-model is not a
definitive  solution, at least it shows that the baryonic  mean mass density could 
play an unexpected  role in the determination of the galactic  rotational
velocities and that both are strongly correlated. If this model is on the right
track, then the rotational velocities alone could allow us to directly determine the 
baryonic mean mass density (and not indirectly from the brightness measurements)  and vice versa in a self-consistent manner. In this case, the
delicate step, i.e.  brightness  $\rightarrow{}$ mean mass density  would be short-circuited.
 After an analysis of spiral galaxies and galaxy clusters,  the work is far from finished.  The $\kappa$-model has to be still applied to the elliptical galaxies, to the globular clusters, to the formation and  stability  of primordial galaxies,   and eventually  to   CMB/cosmology.  Let us also  note  the very captivating open   debate concerning the wide binary stars ([54] versus [55]).  The $\kappa$-model obviously predicts a very weak  $\kappa$-effect  in the immediate vicinity of the Sun, i.e. the motion  of the wide binaries is predicted to be quasi-keplerian in this region.  Much work remains to be done. It would  be  interesting  to concomitantly  perform  the same  studies, on the same collection of galaxies, with  other   models, such as the Newtonian Fractional-Dimension Gravity [29, 30, 31, 32], the   Refracted Gravity [33, 34] and also  the CFT model [20].

\begin{appendices}

\section{Section Rotation curve fits results}\label{secA}

\begin{center}
\includegraphics[height=430pt, width=410pt]{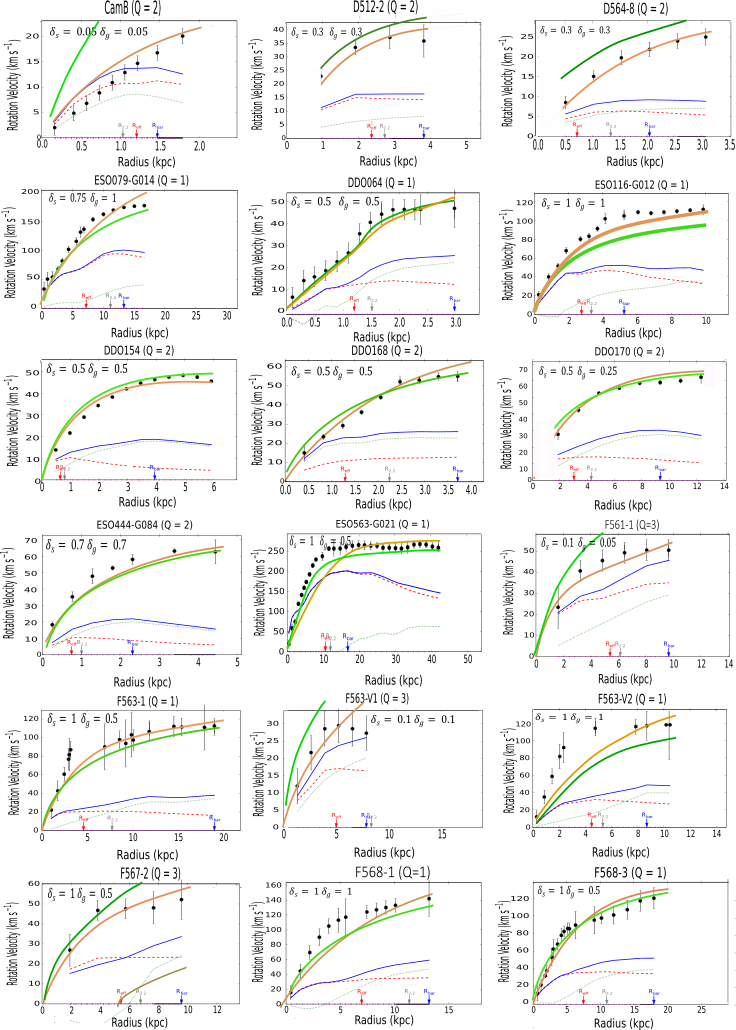}
\end{center}

Figure A Rotation curves of the SPARC galaxies. The green  line  is
predicted by MOND, the amber line is predicted by the $\kappa$-model, the red
dashed  line represents the stars,  the green dotted line represents the gas, the
blue line represents the sum of all baryonic components (stars + gas). The
observed velocities are shown as a  series of filled black circles.

\begin{center}
\includegraphics[height=430pt, width=410pt]{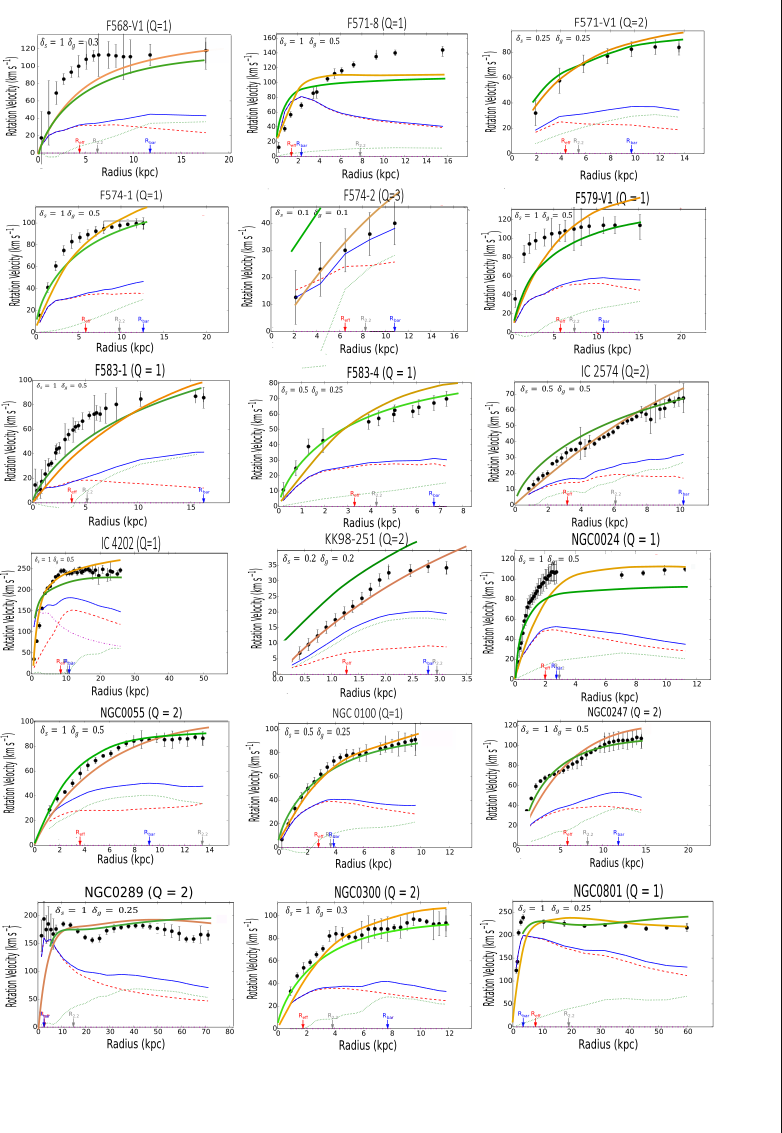}

Figure A :  Continued rotation profiles
\end{center}

\begin{center}
\includegraphics[height=430pt, width=410pt]{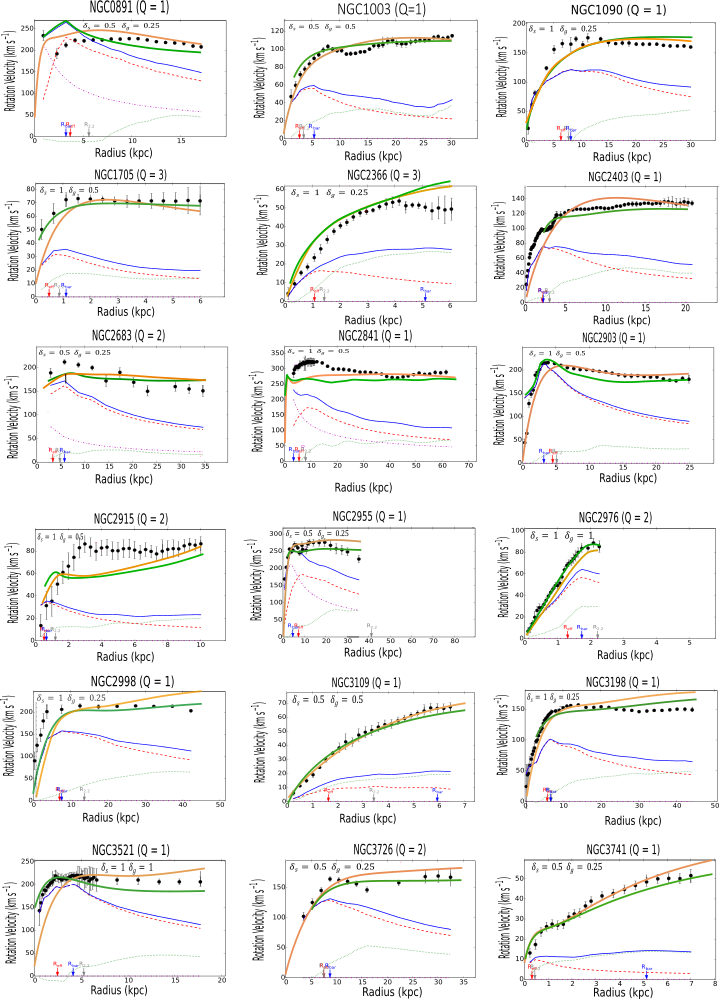}

Figure A :  Continued rotation profiles
\end{center}

\begin{center}
\includegraphics[height=430pt, width=410pt]{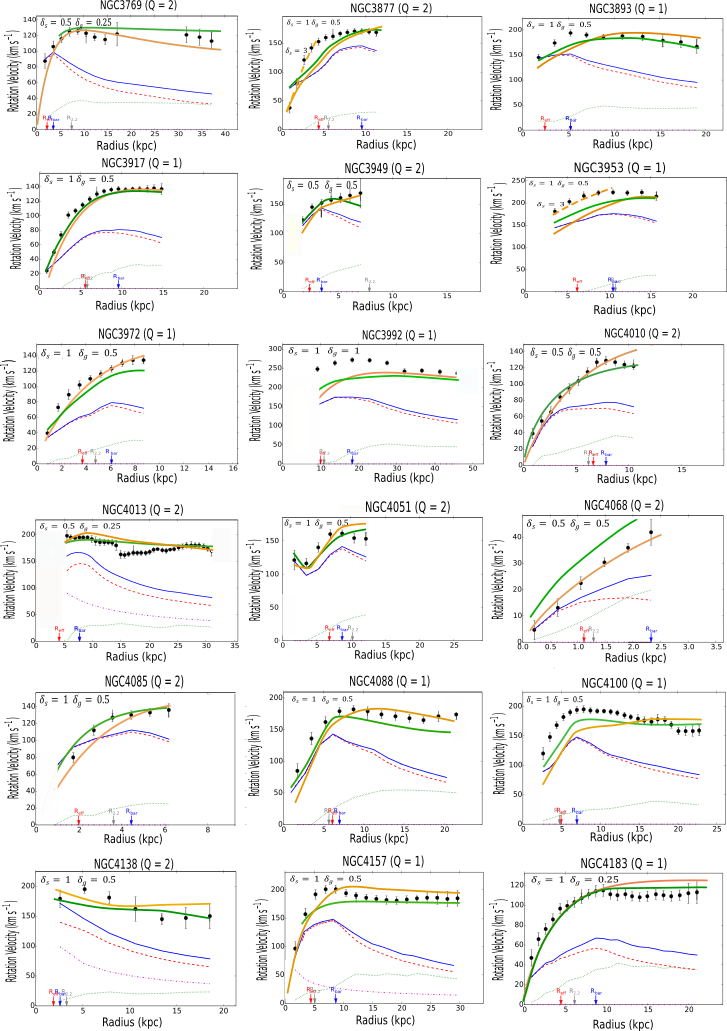}

Figure A :  Continued rotation profiles
\end{center}

\begin{center}
\includegraphics[height=430pt, width=410pt]{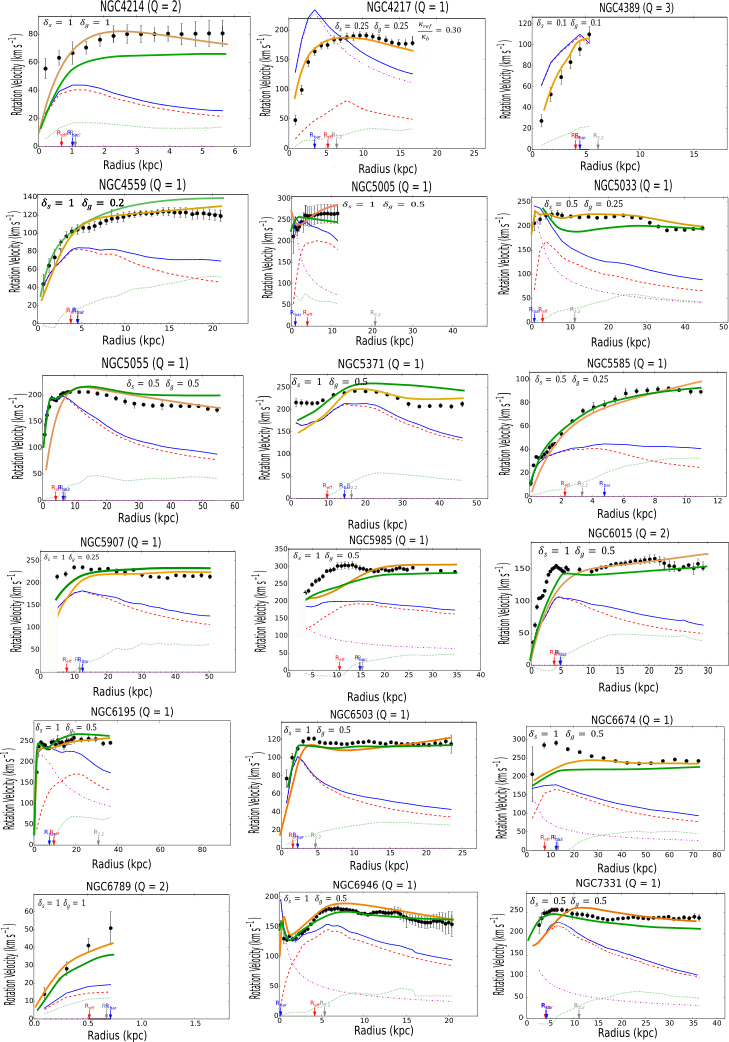}

Figure A :  Continued rotation profiles
\end{center}

\begin{center}
\includegraphics[height=430pt, width=410pt]{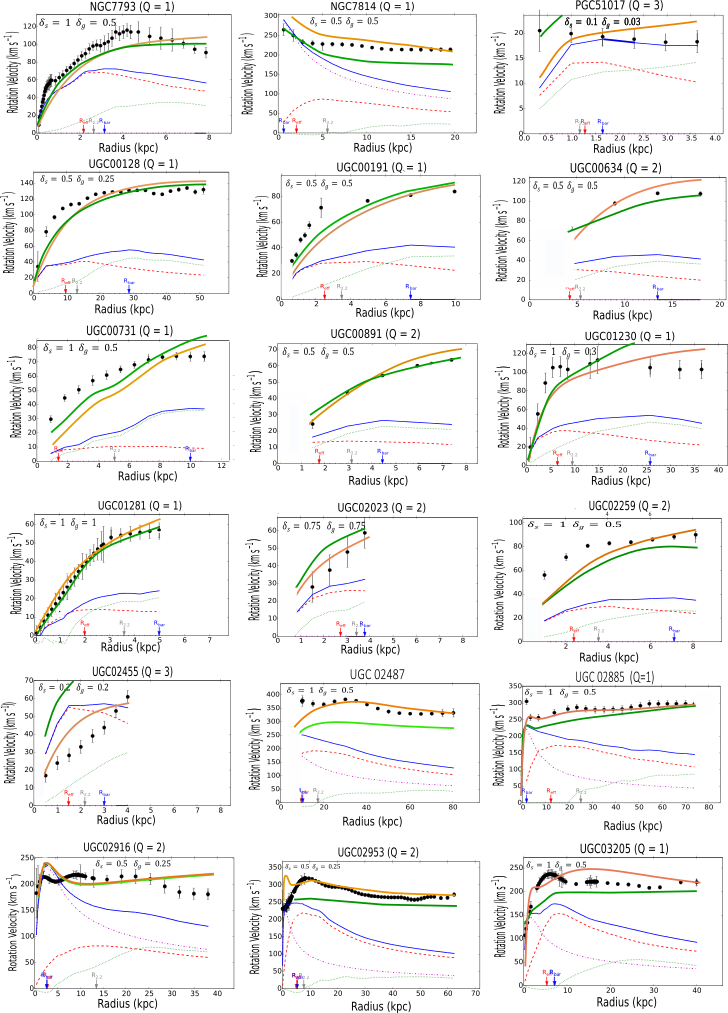}

Figure A : Continued rotation profiles
\end{center}

\begin{center}
\includegraphics[height=430pt, width=410pt]{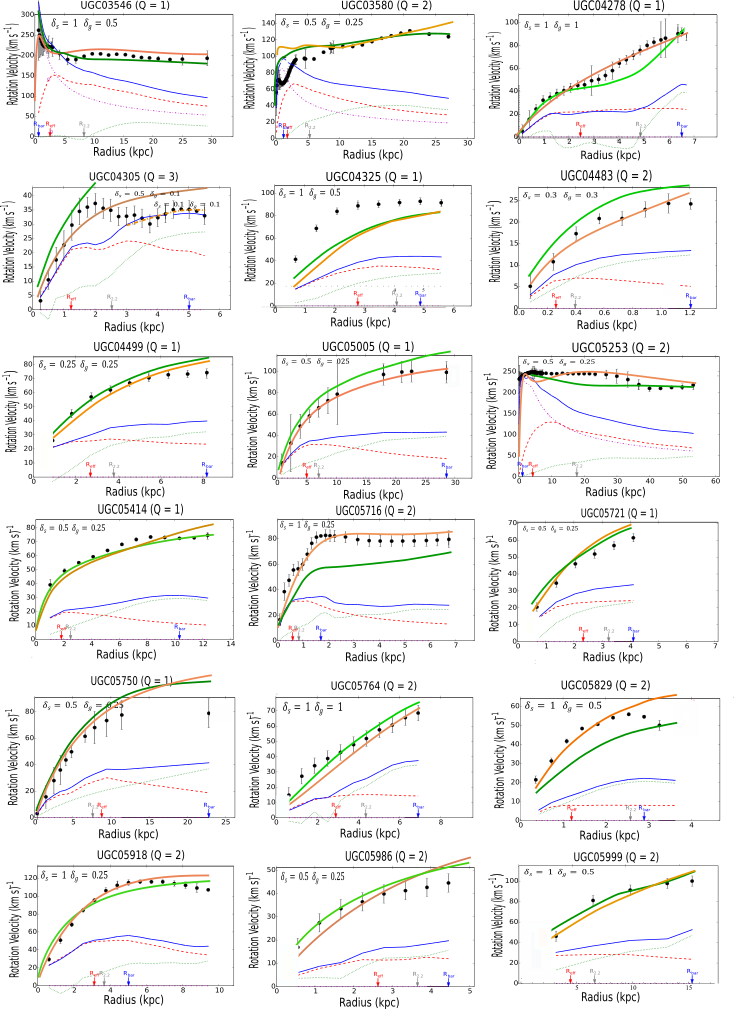}

Figure A :  Continued rotation profiles
\end{center}

\begin{center}
\includegraphics[height=430pt, width=410pt]{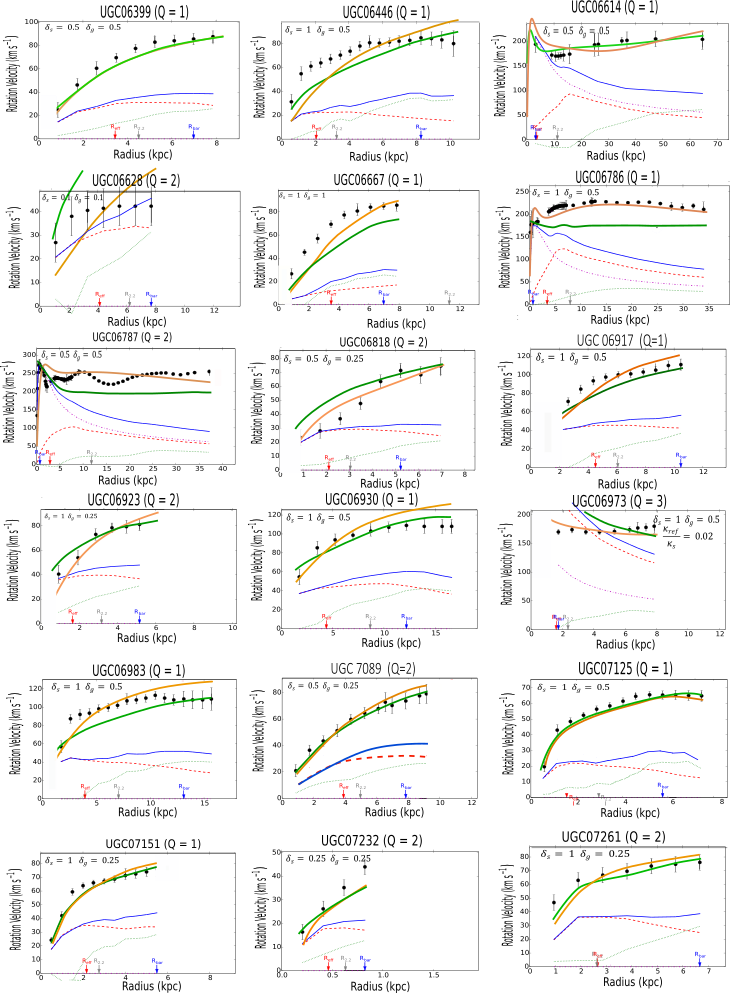}

Figure A  : Continued rotation profiles
\end{center}

\begin{center}
\includegraphics[height=480pt, width=410pt]{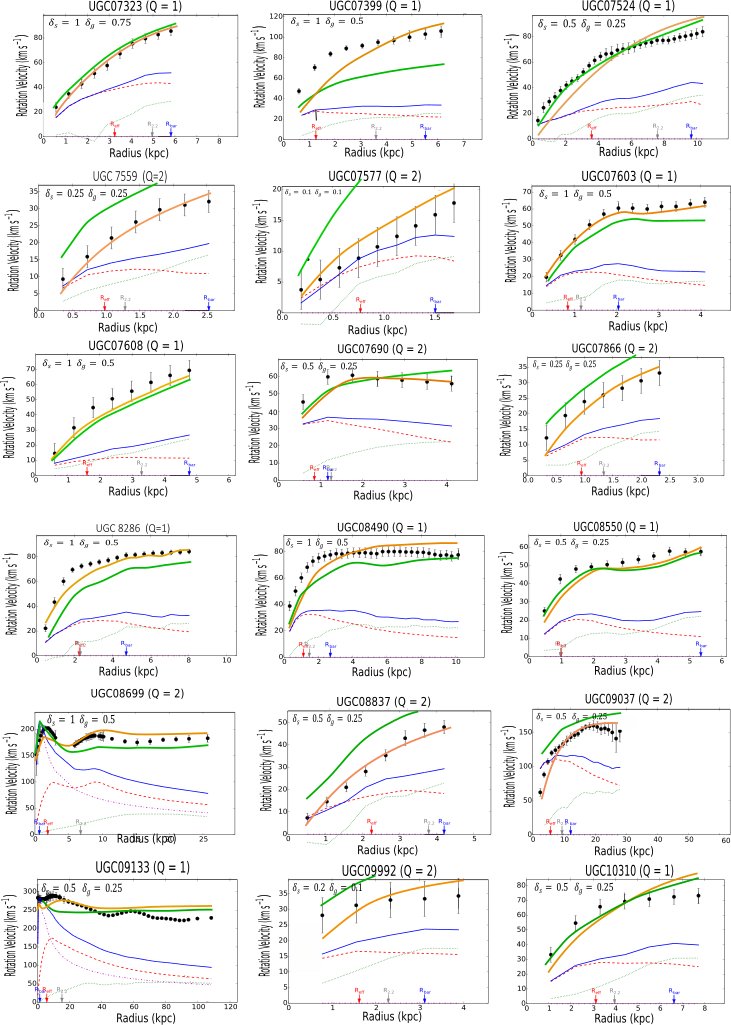}

Figure A  : Continued rotation profiles
\end{center}

\begin{center}
\includegraphics[height=170pt, width=410pt]{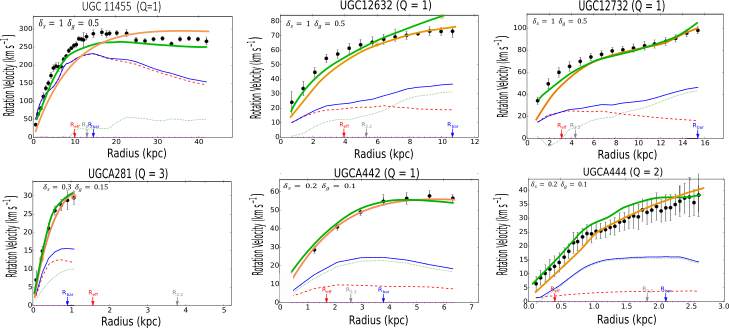}

Figure A  : Continued rotation profiles
\end{center}

\end{appendices} 

\vspace{10pt}
{\raggedright{\textbf{ Acknowledgments}: The author  sincerely thanks the three  referees whose suggestions and corrections improved
this paper.}

\vspace{10pt}
{\raggedright{\textbf{Data availability statement}: The author confirms that the data supporting the findings of this study are available within the article and the reference list.}}

\vspace{10pt}
{\raggedright
{\textbf{Conflicts of Interest:} The author declares no conflict of interest.}}

\vspace{10pt}

{\raggedright
{\textbf{References}
}

[1] Mirhosseini A. and  Moniez M.,  A\&A, 2018, 618, L4

[2] Ackermann M., et al.,  Phys. Rev. Lett., 2015, 115, 231301

[3] Battaglieri M. et al, 2017, arXiv:1707.04591 [hep-ph]

[4] Amaudruz P A et al., (DEAP Collaboration)  Astropart. Phys., 2019, 108, 1

[5] Amaudruz P A et al., (DEAP Collaboration)  Nucl. Instrum. Methods Phys.Res. A, 2019, 922, 373

[6] Arnquist I. et al (Damic-M Collaboration),   Physical Review Letters, 2023, 130, 171003

[7] McGaugh S.,  Can. J.  Phys., 2015, 93, 250

[8] Di Paolo C. Salucci P.  and  Erkurt A.,  MNRAS, 2019,  490, 5451 

[9] Milgrom M.,  Astrophysical Journal, 1983, 270, 365

[10] Milgrom M.,  Can. J. Phys.,  2015, 93(2), 107 

[11] Milgrom  M., Studies in History and Philosophy of Modern Physics, 2020, 7,170

[12] McGaugh S., Astrophys. J., 2005, 632, 859

[13] Famey B. and Mc Gaugh S., Living Rev. Relativity, 2012, 15, 10 

[14] Hodson A.O. and Zhao H., A\&A, 2017, 598, A127

[15] Skordis C, and  Z\l{}o\'{s}nik T.,   Phys. Rev. Lett., 2021, 127, 161302

[16] Pascoli G., Can. J. Phys.,  2023,   101, 11

[17] Moffat, J.W.,  JCAP, 2006, 2006, 004

[18] Rouhani S., and S. Rahvar S.,  MNRAS, 2024, 427, 2831 

[19] Mannheim P.D., Prog.Part.Nucl.Phys. 2006, 56, 340

[20] Li Q., and Modesto L., Grav.  Cosmol., 2020, 26, 99

[21] Ghosh S.,  Bhattacharya M.,  Sherpa1 Y., and  Bhadra A.,  JCAP, 2023, 07, 008

[22] Verlinde E.,  SciPost Phys., 2017,   2, 016

[23] Varieschi G., Found Phys., 2020,  50, 1608

[24] Cesare V., Diaferio A., Matsakos T., and Angus, G. 2020, A\&A, 637, A70

[25] Pascoli, G., and Pernas, L., 2020, hal.archives-ouvertes.fr/hal-02530737

[26] Pascoli G., Astrophys. Space Sc., 2022, 367, 62

[27] Pascoli G., arXiv:2307.01555 [astro-ph.CO]

[28] Calcagni G., JCAP, 2013, 12, 041

[29] Varieschi G., Eur. Phys. J. Plus, 2021, 136, 183

 [30]  {
 { Varieschi G., 2021, MNRAS. 503  2, 1915}}

[31] Varieschi G., Eur. Phys. J. Plus, 2022, 137, 1217

[32] Varieschi G., Universe, 2023, 9(6), 246

[33] Cesare V., Diaferio A., and Matsakos T., 2022, A\&A, 657, A133

[34] Cesare V., Universe, 2023, 9, 56

[35]   Lelli F.,  McGaugh S.S., and  Schombert J.M., Astron. J., 2016, 152,157

[36] {
{ Valenti, E.,  Zoccali, M.,   Mucciarelli, A.,  Gonzalez, O.A.,  Surot1, F.,  Minniti, D.,  Rejkuba1, M.,  Pasquini, L.,  Fiorentino, G.,  Bono, G.,  Rich, R.M., and  Soto, M., A\&A, 2018, 616, A83}}

 [37]  Ogle P. M., Jarrett T., Lanz L., Cluver M., Alatalo K., Appleton P.N. and Mazzarella J. M.,  ApJ. Letters, 2019, 884

[38] {
{ Lisenfeld, U.,   Ogle, P.M.,  Appleton, P.N.,  Jarrett, T.H., and Moncada-Cuadri, B.M., 2023, A\&A,  673, A87}}

[39]  de Vaucouleurs, G. 1958, ApJ, 128, 465

[40] Yoachim P. and Dalcanton J..J, AJ,  2006, 131, 226

[41]   Johnson M.C., Hunter D.A., Kamphuis P., and Wang J. MNRAS, 2017, 465, L49

[42]   Li X., Shi Y., Zhang Z.Y., Chen J., Yu X., Wang J., Gu Q., and Li S., MNRAS, 2022, 516, 4220

[43] {
{ Merritt, D., Graham, A. W.,  Moore, B., Diemand, J. and Terzić, B., AJ, 2006,   132,  6,  2685}}

[44]   Li P., Thesis, Case Reserve Western University, 2020

[45]  Ghari A., Famaey B.,  Laporte C., and Haghi H., A\&A, 2019, 623, A123

[46]   Sofue Y. 2015 PASJ, 68, 2

[47]   Mc Gaugh S.S.,  ApJ, 2019, 885, 87

[48]  Mancera Pi\~{n}a P.E., Fraternali F., Oosterloo T., Adams E.A.K., Oman K.A., and Leisman L., MNRAS, 2022, 512, 3230

[49]   Mc Gaugh S., Studies in History and Philosophy of Science, 2021, 88 suppl C, 220 https://doi.org/10.1016/j.shpsa.2021.05.008

[50]   Loizeau N. and  Farrar G.R., ApJ Letters, 2021, 920, L10

[51]  Randriamampandry T.H and  Carignan C.,  MNRAS, 2014, 439, 2132

[52]  Haghi H.,  Bazkiaei A.E.,   Zonoozi A.H., and  Kroupa P., MNRAS, 2017, 469, 3909 

[53]   Nieuwenhuizen T. M., 2017, Fortschr. Phys., 65, 201600050

[54]  Hernandez X.,  Verteletskyi V., Nasser L., and  Aguayo-Ortiz A.,  arXiv:2309.10995v2

[55]   Banik I., Pittordis C., Sutherland W., Famaey B., Ibata R.,  Mieske S., and Zhao H., arXiv:2311.03436v1

\end{document}